\documentclass{article}

\usepackage{arxiv}

\usepackage[utf8]{inputenc} 
\usepackage[T1]{fontenc}    
\usepackage{hyperref}       
\usepackage{url}            
\usepackage{booktabs}       
\usepackage{microtype}      
\usepackage{graphicx}
\usepackage{natbib}
\usepackage{amsmath}

\graphicspath{ {./images/} }

\begin{document}

\title{Non-Stationary Large-Scale Statistics of Precipitation Extremes in Central Europe}

\author{
 Felix S. Fauer\\
  Institute of Meteorology\\
  Freie Universität Berlin\\
  Carl-Heinrich-Becker Weg 6-10, 12165 Berlin, Germany\\
  \texttt{felix.fauer@met.fu-berlin.de} \\
   \And
 Henning W. Rust\\
  Institute of Meteorology\\
  Freie Universität Berlin\\
  Carl-Heinrich-Becker Weg 6-10, 12165 Berlin, Germany\\
  \texttt{henning.rust@fu-berlin.de} \\
}

\maketitle

\begin{abstract}
Extreme precipitation shows non-stationary behavior over time, but also with respect to other large-scale variables. While this effect is often neglected, we propose a model including the influence of North Atlantic Oscillation, time, surface temperature and a blocking index. The model features flexibility to use annual maxima as well as seasonal maxima to be fitted in a generalized extreme value setting. To further increase the efficiency of data usage maxima from different accumulation durations are aggregated so that information for extremes on different time scales can be provided. Our model is trained to individual station data with temporal resolutions ranging from one minute to one day across Germany. The models are selected with a stepwise BIC model selection and verified with a cross-validated quantile skill index. The verification shows that the new model performs better than a reference model without large scale information. Also, the new model enables insights into the effect of large scale variables on extreme precipitation. Results suggest that  the probability of extreme precipitation increases with time since 1950 in all seasons. High probabilities of extremes are positively correlated with blocking situations in summer and with temperature in winter. However, they are negatively correlated with blocking situations in winter and temperature in summer.
\end{abstract}

\section{Introduction}\label{sec1}

Hydrologic extremes are changing. This is supported by the sixth IPCC assessment report (AR6) \citep{ipcc6_cha11} which finds that the majority of measurement stations in Europe showed a significant increase in extreme precipitation over durations of 1 day and 5 days between 1950 and 2018. Trends might be variable in sign and value across regions and seasons \citep{croitoru2013, fischer2015, chiew2009, arnbjerg2012}. 
For example, a decreasing 5-day-maximum-precipitation (RX5day) by the year 2100 is reported \citep{ipccAtlas_a, ipccAtlas_b} in a 2$^{\circ}$C warming scenario in summer and increasing RX5day in the other seasons. These facts show the heterogeneity of developments in extreme precipitation. Furthermore, they emphasize that precipitation extremes are changing in a non-stationary fashion.

Describing extremes and predicting annual probabilities is done with extreme value statistics, where the generalized extreme value (GEV) distribution is fitted to block maxima. Here, one block is considered (1) as a year for annual models or (2) as a month for a seasonal models. Even in a stationary setting, extremes are difficult to model since they are rare by definition. Further increasing the complexity of the model by analyzing the dependence of extremes on other variables is a challenge which can be faced by more efficient usage of data.   Effects occur over different time scales and therefore, precipitation data from different measurement resolutions (from minutes to days) can be accumulated (duration steps). With this data, duration-dependent GEV (d-GEV) distributions \citep{nguyen1998, menabde1999} have been used so that more information of each year is processed, as maxima of different duration steps are fed into the model. The results of such analyses are often shown in Intensity-Duration-Frequency (IDF) curves \citep{chow1953}. The relation between duration and intensity can be described by different parametrizations, including multiscaling \citep{gupta1990}, duration-offset \citep{koutsoyiannis1998} and intensity-offset \citep{fauer2021}. Some of these duration-dependent approaches have been combined with large scale influence on precipitation by \cite{ouarda2019}. In their study, the d-GEV parameters depended on large scale covariates, e.g., time and several teleconnection patterns. This resulted in statistics for three locations in the USA. However, there is no study known to us which covers Central Europe with such a model. Our approach uses a similar method as \cite{ouarda2019} and the main new aspects are: (1) We cover the region of Germany with 199 sites of data. (2) We use other large-scale variables (NAO, blocking, temperature) which might fit better to the atmospheric circumstances in Central Europe. (3) Our model features advanced flexibility regarding different durations and enables more potential dependencies with large scale variables. (4) We use an advanced verification method to assess whether the use of large-scale information improves the model, aside from new insights into large-scale effects.

Our analysis aims for the identification of meaningful large-scale variables. Therefore, we investigate the influence of blockings, North Atlantic Oscillation (NAO), temperature and time.

A blocking situation is characterized by an interuption of the westerly wind flow due to persistent anticyclones \citep{otero2022}. The presence of a blocking situation can influence the appearance of heavy precipitation. The change of odds for heavy precipitation in presence of blocking depends heavily on season and region \citep{lenggenhager2019}. We will compare our findings with the literature with respect to our definition of blocking and choice of region in Sect. \ref{sec:discussion}.

The NAO is the most important teleconnection pattern in Europe \citep{barnston1987}. The change of extreme precipitation with respect to NAO has been investigated by \cite{casanueva2014} and the association between both variables is opposite in winter (positive) and summer (negative) in Germany. There, precipitation trend maps over time in Germany mostly show non-significant coefficients both in summer and winter.

Temperature and extreme precipitation show a correlation which has received considerable attention in the literature \citep{aleshina2021, westra2014}. The Clausius-Clapeyron scaling describes the correlation between potential water content and air temperature and explains why average rain amounts increase in warmer air. However, the connection between extreme precipitation and temperature is more complex. After correcting for the Clausius-Clapeyron scaling, the sign of the correlation coefficient changes depending on the temperature regime and is negative (positive) for warmer (colder) temperatures in Australia \citep{jones2010}. The same applies to Europe, where temperatures above 15$^{\circ}$C lead to less extreme precipitation \citep{drobinski2016}. In North-America, the correlation between both quantities is consistently positive \citep{mishra2012}, which is also known as Clausius-Clapeyron (C-C) scaling.

The temporal trend, described as the change with respect to time as a large scale variable, has to be treated carefully as time is only a proxy for other effects that influence meteorological extremes. On the other hand, those effects are highly non-linear and therefore hard to describe. This is a pro-argument for time as covariate since it might combine unknown effects of other variables.

\section{Data and Methods}\label{methods}

\subsection{Data}

\begin{figure}[h]%
\centering
\includegraphics[width=12cm]{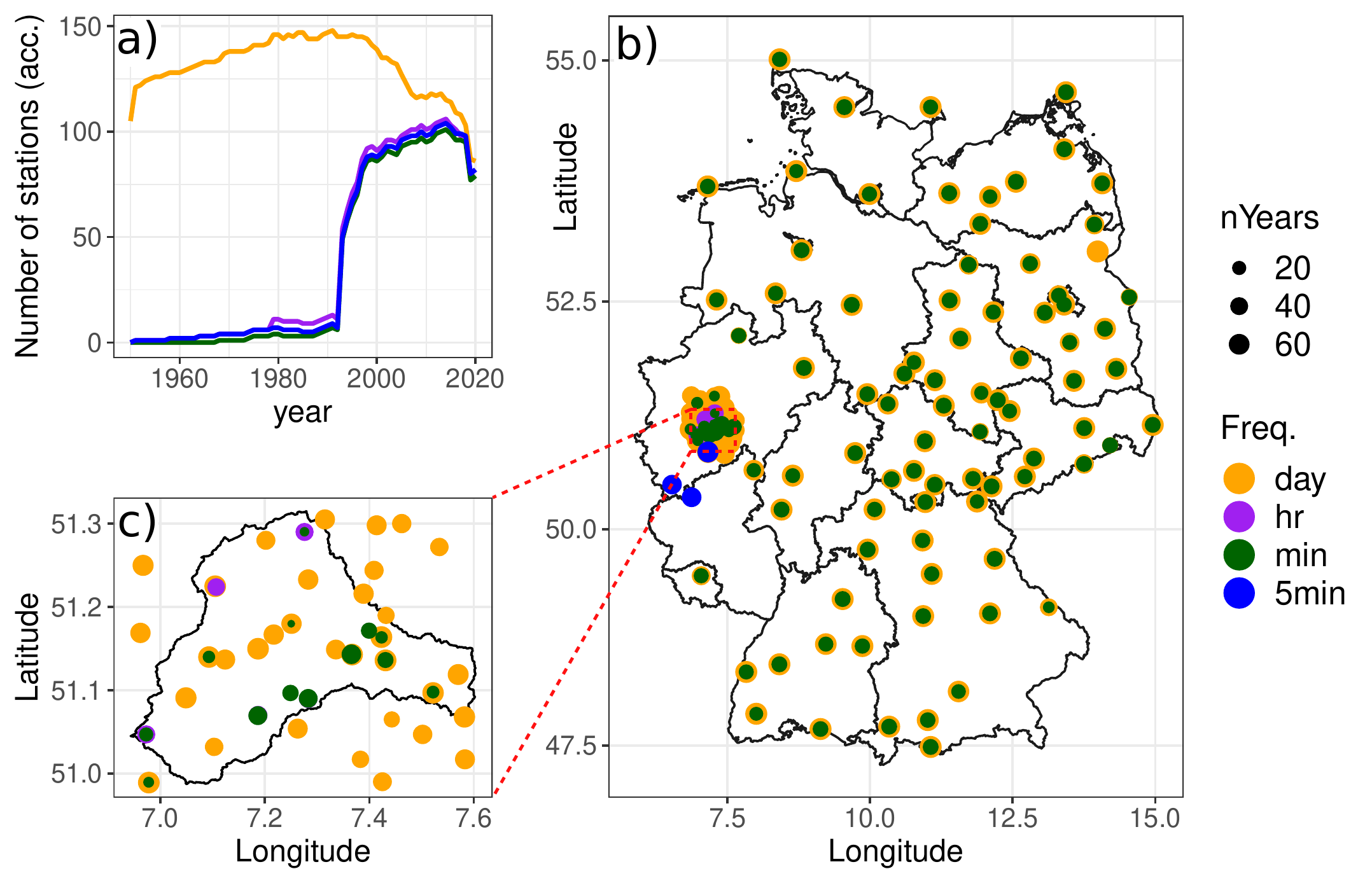}
\caption{
\textbf{a}: Number of stations (accumulated) that provide data of each measurement frequency (color). \textbf{b and c}: Stations (dots) showing measurement frequency (color) and length of time series (radius). \textbf{c}: Zoom into Wupper Catchment area.
}\label{fig:data}
\end{figure}

We use precipitation data from different sources. The German meteorological service (DWD) provides data from 86 stations across Germany that cover both daily and minutely resolution (see Fig. \ref{fig:data}b). This data is publicly available \citep{dwdOpenServer}. Additionally, data from three DWD stations with long time ranges (longest with 57 years) and 5-minute resolutions were provided to us which are not publicly available (see Acknowledgements). Furthermore, the Wupperverband provided data from 57 stations with daily data, 6 stations with hourly data and 18 stations with minutely data (see Fig. \ref{fig:data}c). Stations vary in length of time series and availability of high-temporal-resolution measurements (Fig. \ref{fig:data}a-c). Different stations that have a distance of less than 250m were grouped together, since precipitation amount should not change considerably. Possible duplicates, i.e., more than one value for a specific station and duration and year might occur because different stations were merged or because both minutely and daily measuring devices will provide an accumulated rainfall value for durations $d\geq 24\,h$. In this case, values from the lower measuring frequency are omitted.

The data of the NAO index is obtained from the National Oceanic and Atmospheric Administration (NOAA) and the Climate Prediction Center (CPC), where it is openly available \citep{noaa_nao}. We use the dataset with monthly values which is based on a Rotated Principal Component Analysis, starting in 1950.

The mean surface air temperature (tas) over Germany is obtained from the ERA5 dataset with a daily resolution of 0.25$^{\circ}$.The data is spatially averaged between 4$^{\circ}$W and 15$^{\circ}$W longitude and between 45$^{\circ}$N and 55$^{\circ}$N latitude. This way, one value per time step indicates the mean temperature on a large scale. Data is available from 1950 to 2021 \citep{era5}.

The blocking information is inferred from a binary blocking-index (BBI), using gridded daily ERA5 data (by 2.5$^{\circ}$). It is based on the two-dimensional blocking index from \citep{scherrer2006, schuster2019} with minor modifications. The BBI of the grid fields is averaged over Scandinavia, because atmospheric blocking situations over this region are found to have an influence on convection in Central Europe \citep{mohr2019}. The blocking value that is used here ranges between 0 and 1 and indicates the spatial fraction of grid fields that were identified as blocked.

All daily values of the large-scale variables, i.e., NAO, temperature and blocking, are averaged over non-overlapping blocks of one month or one year, depending on the model (season or annual). Since all datasets of large-scale variables start in 1950, precipitation data of earlier years are omitted because our model cannot handle missing values in any of the predictor terms.

\subsection{Methods}

Extreme precipitation is commonly modeled with the block-maxima approach and the GEV distribution. This distribution links probabilities or return periods and return levels. In this study, an extended version of the d-GEV distribution is used as proposed by \cite{fauer2021} which has a higher flexibility for very short ($d < 8\,$h) 
and very long ($ d > 24\,$h) durations. This flexibility is introduced by a combination of existing features, namely curvature for short durations and multiscaling for medium durations, and an extension with an additional parameter $\tau$ which allows for return levels to deviate from the log-linear relation to duration \citep{fauer2021, ulrich2021}. This flexible model is described by

\begin{align}
	G(z) &= \exp \bigg\{ -\left[1+\xi \left( \frac{z-\mu(d)}{\sigma(d)}\right)\right]^{-1/\xi} \bigg\}, \label{eq:dGEV} \\	
  \sigma(d)&=\sigma_0 (d+\theta)^{-(\eta + \eta_2)} + \tau,\\
  \mu(d) &= \tilde{\mu} (\sigma_0 (d+\theta)^{-\eta} + \tau),
\end{align}

with the rescaled location parameter $\tilde{\mu}$, the scale offset $\sigma_0>0$, the shape parameter $\xi \neq 0$, the duration offset $\theta>0$, the two duration exponents $\eta$ and $\eta_2$, the intensity offset $\tau>0$ and duration $d>0$. The intensity $z$ is restricted to $1+\xi(z-\mu)/{\sigma_0}>0$. If $\xi=0$, then $G(z)=\exp\{-\left[\exp((z-\mu(d))/\sigma(d))\right]\}$ applies.

Using the annual maxima, i.e., one value per year, easily enables the estimation of average return periods $T$ since it is connected to the annual non-exceedance probability $p$ from the GEV distribution function by $T = 1/(1-p)$. Consequently, the exceedance probability is $p_e =1-p$. When using monthly maxima with values for a whole season of 3 months, i.e., 3 values per year, the probability $p_s$ from the distribution function has to be converted with $p = 1-(1-p_s)^{1/3}$ to get annual non-exceedance probabilities $p$, again.

The distribution is fitted to the data with a maximum likelihood estimation (MLE), meaning that the distribution parameters are chosen in a way such that the combined probability of all data points is maximized. Multiplying many probabilities (numbers between 0 and 1) might lead to numerical instability. That is why we use the log-likelihood (LL) which does not require a multiplication but sum of log probabilities 

\begin{align}
LL=\log(\prod_i^n p_i)=\sum_i^n \log(p_i)
\end{align}

with $p_i$ being the estimated probability of the data point with index $i$ and $n$ the number of data points.

The uncertainty of estimated intensities is obtained by parametric bootstrapping in which data points are sampled 1000 times and the distribution is modeled for each bootstrap. Then, the 2.5th and 97.5th percentiles of the bootstrapped intensity estimates are taken as confidence interval.

\subsection{Non-stationarity in d-GEV parameters}

The need to incorporate non-stationarity into models of precipitation extremes will be shown in this section with sliding window models. Here, data points are grouped according to the value of a large scale variable and a model is trained to each group. This way, the change of d-GEV parameters can be shown with respect to a large scale variable. The results of this section do not have a numerical impact on the model selection, described in Sect. \ref{cha:ms}.

\begin{figure}[h]%
\centering
\includegraphics[width=1\textwidth]{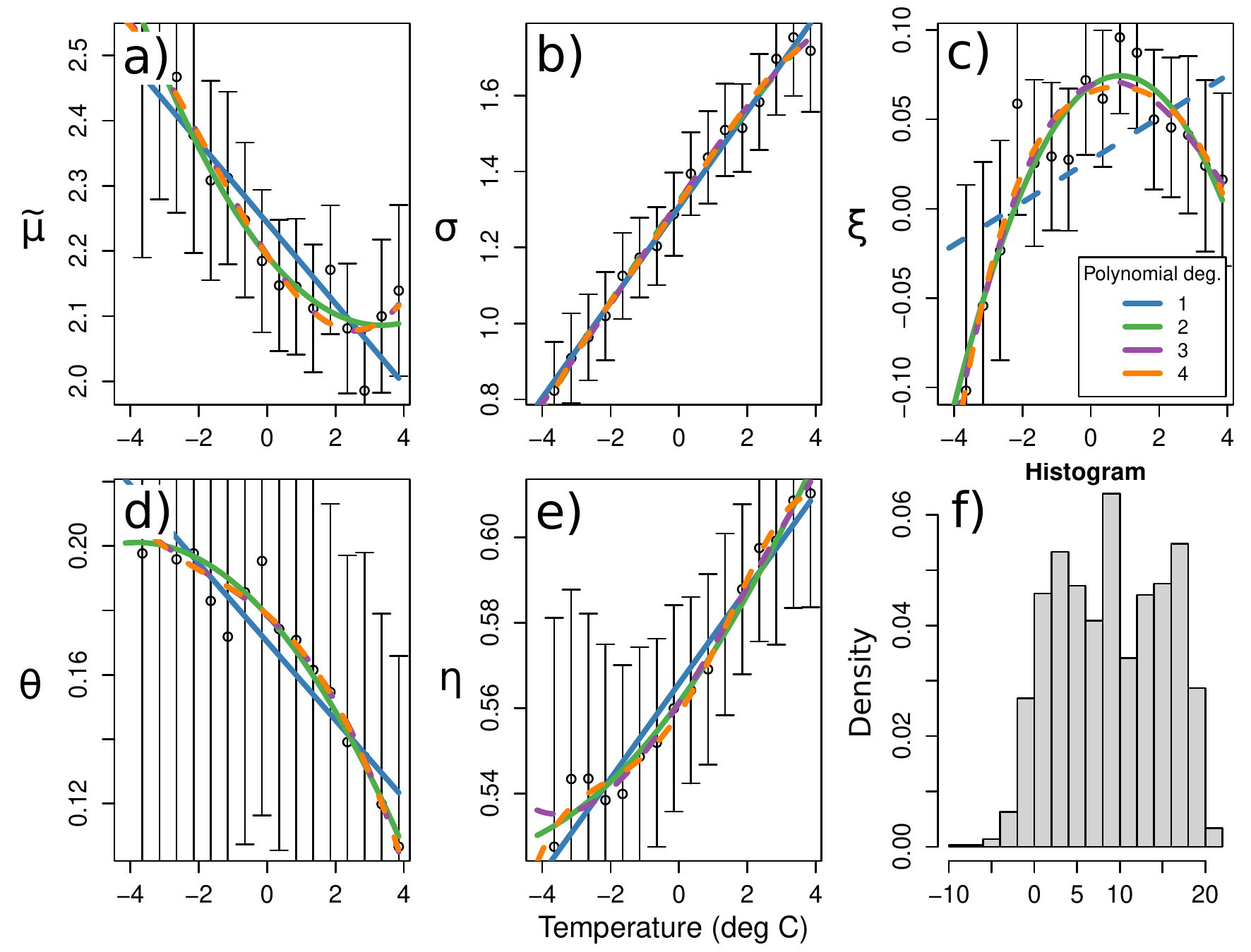}
\caption{Dependence of flexible d-GEV parameters on large-scale variable temperature. \textbf{a-e:} Parameter value over temperature. Solid lines represent polynomials of significant least-squares fits according to a t-test. Data from an example station Nürburg-Bahrweiler, for winter. \textbf{f:} Histogram of temperature data from all stations.
}\label{fig:params_dependence}
\end{figure}

The dependence of d-GEV parameters $\tilde{\mu}$, $\sigma_0$, $\xi$, $\theta$ and $\eta$ on the large-scale variable temperature is shown in Fig. \ref{fig:params_dependence} for a selected station (Nürburg-Barweiler) in winter. This figure illustrates the necessity of incorporating large scale variables in the model for the GEV parameters. Subsets of the data were created by choosing overlapping ranges of 4$\,^{\circ}$ around all possible centered temperature values in the data (first subset: -6$\,^{\circ}$C to -2$\,^{\circ}$C, second subset: -5.5$\,^{\circ}$C to -1.5$\,^{\circ}$C, ...). Then, the model was trained with each of these subsets. The model parameters depending on the chosen subset with the centered temperature value on the x-axis are plotted as dots with vertical uncertainty bars. The uncertainty is the standard error of the maximum likelihood estimate for this parameter.
The four lines in different color represent a least-squares fit to polynomials of degree 1 to 4. Solid lines indicate a significant fit of the covariate with the highest polynomial order according to a two-sided t-test with a p-value below 0.025. Dashed lines indicate a non-significant fit. For example, the rescaled location parameter $\tilde{\mu}$ shows a significant dependence of temperature with a polynomial of orders 1 or 2 (blue and green solid lines). The rescaled location parameter $\sigma_0$ shows a linear trend (solid blue line). And the shape parameter $\xi$ has a significant dependence of polynomial degree 2 (solid green line). The histogram (Fig. \ref{fig:params_dependence}f) shows that temperature values from all stations are not uniformly distributed and explains the higher uncertainty for very low temperatures.

Based on the results of this section, the boundaries for the model selection were set, i.e., which d-GEV parameters will be used for each season, and which large-scale variables will potentially be used for estimating the d-GEV parameters. The complex model with seven parameters is not used in all cases. For stations without sub-hourly data, we do not use the flexible IDF-model, because the complexer models are not expected to improve results \citep{fauer2021}. Here, only the parameters $\tilde{\mu}$, $\sigma_0$, $\xi$ and $\eta$ are used, the others are set to zero. For winter models (DJF), the parameter $\tau$ was set to zero because the purpose of this parameter was to model annual maxima despite the extremes of different duration regimes happening in certain seasons \citep{fauer2021, ulrich2021}. In summer models (JJA) this parameter seems to improve the model since trends of $\tau$ over large-scale variables were often significant and therefore we did not set it to zero. Alternatively, all possible combinations could have been tested and the model selection would have had the possibility to decide on the usage of all parameters. But, these limitations reduce the computing costs of the following analysis and can partially be justified theoretically.

In the final model, each of the chosen d-GEV parameters can potentially depend on each of the large-scale variables up to a polynomial of fourth order. Equations that describe the final models of selected stations are shown in appendix section \ref{cha:apx_ms}. The maximum number of dependencies for each parameter is set to two.

\subsection{Model selection} \label{cha:ms}

\begin{figure}
\centering
\includegraphics[width=1\textwidth]{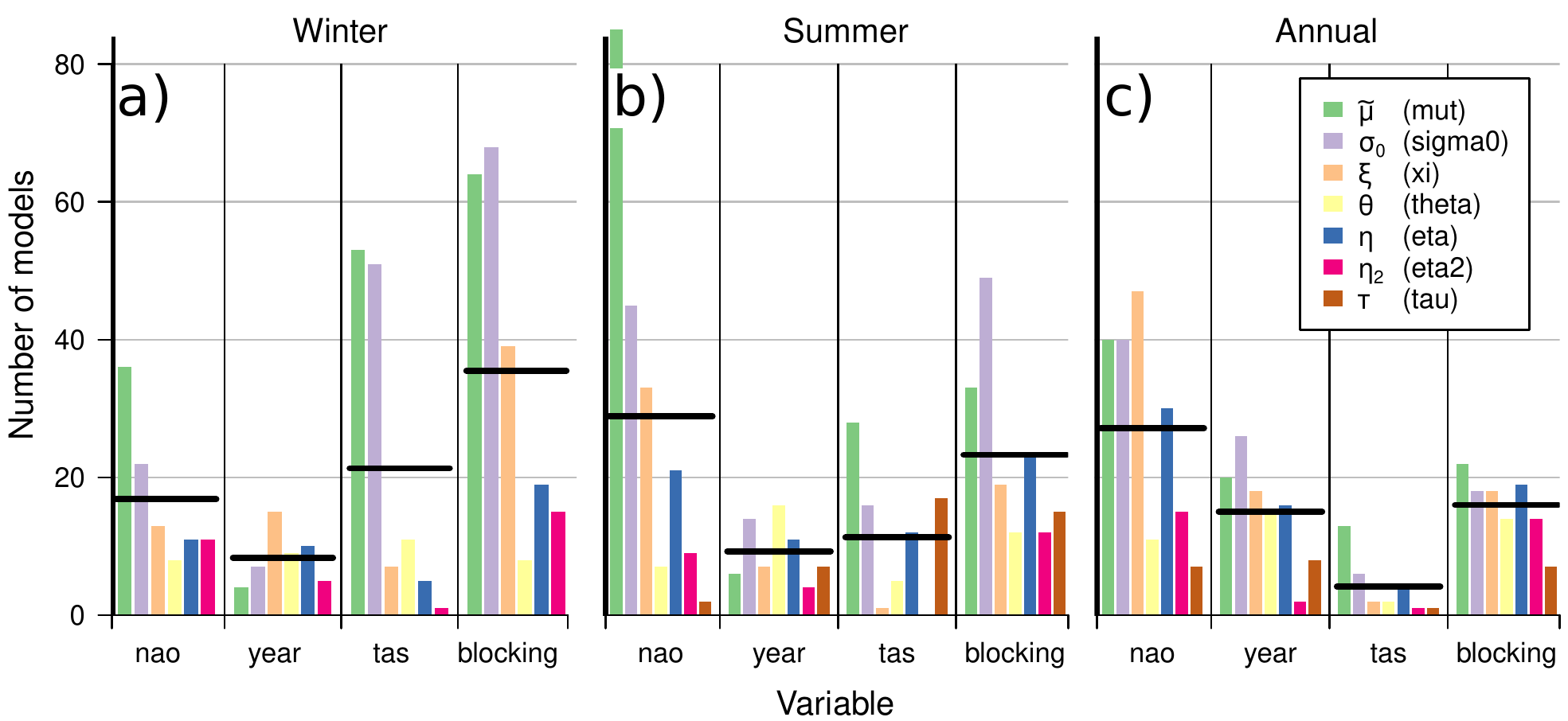}
\caption{For each season (\textbf{a-c}) and variable, the bars show how often a parameter was chosen in a stepwise BIC model selection. Horizontal black lines show the average number of models for this variable and indicate the importance of each variable in the respective season.}\label{fig:ms_stats}
\end{figure}

Not all d-GEV parameters show a significant dependence on large-scale variables and the usage of too many parameters increases their uncertainty \citep{dibaldissarre2006}. Also, overfitting might be a potential problem. Therefore, we conducted a stepwise Bayesian information criterion (BIC) model selection for each station individually as follows: The initial reference model is a d-GEV model without any large-scale dependence. Then, all possible parameter-variable dependencies (combinations) of d-GEV parameters (7), large-scale variables (4) and order of polynomial (4) are added individually (7*4*4=112 possible models) in parallel. Whichever model scores the lowest BIC is selected as the new reference model. Then, again all remaining possible model combinations are added to the new reference model in turns. This procedure is repeated until none of the new models has a lower BIC than the reference model. Figure \ref{fig:ms_stats} shows the number of models in which each parameter-variable combination has been chosen. The black horizontal lines show the mean number of models for this variable. In winter, blocking was chosen more often than other variables. In summer and annual models, NAO has the most influence on the model. It was unexpected that NAO has more influence in summer than in winter. All in all, large-scale dependencies were chosen most often by the location $\tilde{\mu}$ and scaling $\sigma_0$ parameter.

\subsection{Quantile Skill Index}
The new model with large-scale information is compared to a reference model without large-scale information. This is done with the Quantile Skill Index ($-1 \le QSI \le 1$) which is based on the Quantile Skill Score ($QSS\le 1$). The QSS evaluates the quantile score ($QS>0$) \citep{bentzien2014} of a model, compared to the QS of a reference. The QS compares the predicted quantile of a model with all data points and penalizes data points that are higher than the predicted quantile with a weight that scales with the according non-exceedance probability. This way, the model is penalized strongly, when data points exceed predicted quantiles with a high non-exceedance probability. For a given probability $p$ and duration, the QSI shows whether the new model predicts according $p$-quantiles better (values close to 1) than the reference model or worse (values close to -1). The same procedure has been described in detail by \citet[][Section 2.5]{fauer2021}. To prevent overfitting, the QSI is cross-validated (CV) by using every possible three subsequent years as testing set and the remaining years as training set (test set in the first CV step: year 1 to 3, second CV step: year 2 to 4, ...).

\section{Results} \label{sec:results}

\subsection{Verification}

\begin{figure}
\centering
\includegraphics[width=1\textwidth]{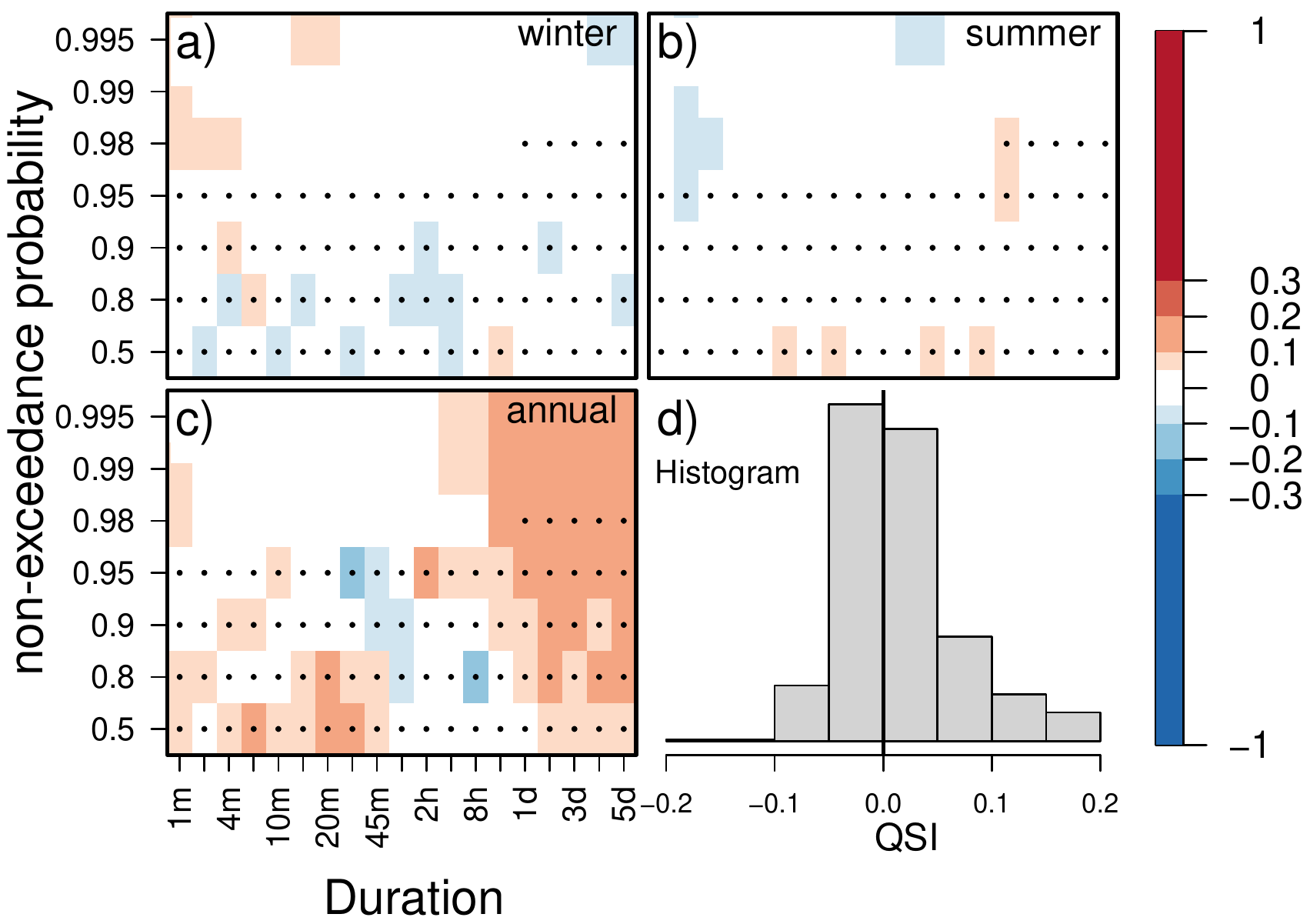}
\caption{Verification of large-scale model. \textbf{a-c:} The QSI is shown for every probability and duration with the stationary model as reference model. Positive values of QSI (red color) indicate an improvement of the large-scale model over the reference model. \textbf{d:} Histogram over all quantile skill indices. Most QSI values are between -0.05 and 0.05 (colored in white in a-c) and are considered non-relevant.
}\label{fig:verification}
\end{figure}

The large-scale flexible d-GEV models were verified against flexible d-GEV models without large-scale dependence using the QSI median over all stations. Figure \ref{fig:verification} shows the QSI for all durations from 1 min to 5 days and probabilities/return periods up to 200 years and all seasons (a-c). Probabilities higher than $p=0.98$ have to be handled with care, because the quantile score cannot reasonably evaluate return periods longer than the time range of the data. This is, because in this regime, a model is incentivized to predict larger values, since all data points are lower than the modeled quantile and the QS penalizes larger data points stronger. Therefore, black dots indicate whether the average number of years is equal or higher than the return period corresponding to the probability (vertical axis).

In annual models (Fig. \ref{fig:verification}c), the large-scale model has a higher QSI in most durations and probabilities while in winter DJF (a) and summer JJA (b), there is no clear tendency. Despite there being no improvement of non-stationary modeling in many duration/probability regimes, the new models gain insight into dependencies (see Sect. \ref{cha:idv}). The color-scale exceeds the range of values in the plot because it is chosen consistently with previous studies evaluating the QSI of d-GEV models \citep{fauer2021, ulrich2020}.

\subsection{Large-scale dependence of extreme precipitation} \label{cha:idv}

\begin{figure}[h]%
\centering
\includegraphics[width=1\textwidth]{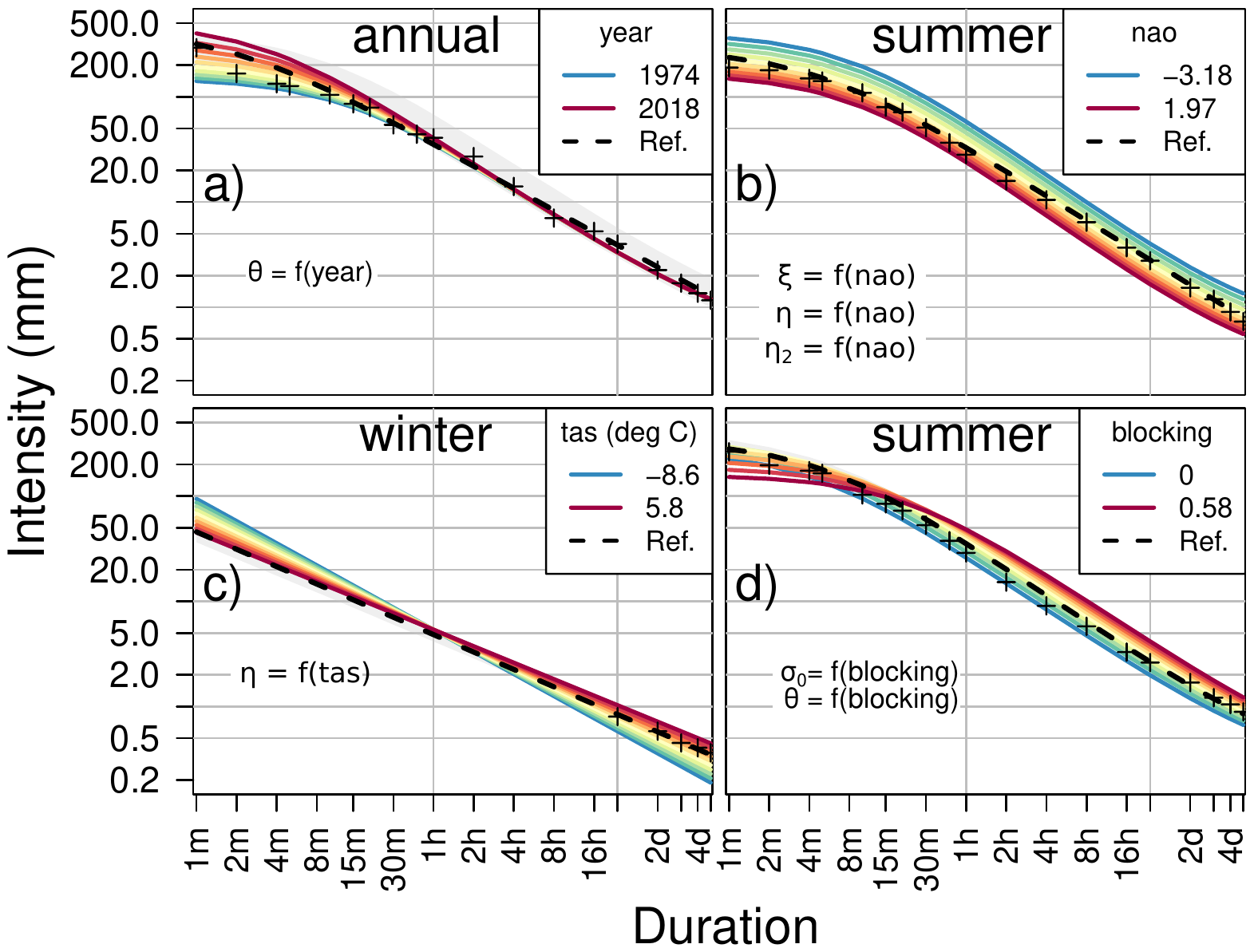}
\caption{Intensity-Duration-Variable (IDV) curve for selected stations. Intensity over duration is shown for different values of large-scale variables. Stations are chosen for the purpose of visualizing the effect on d-GEV parameters. They do not necessarily represent a general trend over all stations. Dashed lines show a reference IDF curve without large-scale modeling. The semi-transparent grey background shows the confidence interval of the reference, obtained by bootstrapping. \textbf{a:} Annual model of station Lindscheid, visualizing the effect on duration offset. \textbf{b:} Summer model of station Uckermünde, visualizing the effect on shape. \textbf{c:} Winter model of station Angermünde, visualizing the effect on duration exponent (only daily data) \textbf{d:} Summer model of station Doberlug-Kirchhain, visualizing the effect on duration offset and scaling parameter.}\label{fig:idv}
\end{figure}

We present a visualization of modeling large-scale precipitation extremes which is an adaptation of known IDF curves (Fig. \ref{fig:idv}). The axes for intensity and duration stay the same, but different curves and colors show the range of a large-scale variable while the exceedance probability (average return period) is fixed to $p_e=0.05$ (20 years) and the other large-scale variables are fixed to an average value. We call this visualization Intensity-Duration-Variable (IDV) curve. A stationary reference model without large-scale training is added (dashed line). 
In a model where the duration offset $\theta$ depends on the year, intensities will vary only for short durations (Fig. \ref{fig:idv}a). Dependence of location $\tilde{\mu}$, scale $\sigma_0$  or shape $\xi$ (Fig. \ref{fig:idv}b and d) will let the intensities vary over the whole range of durations equally (on a log-scale) and produce a shift along the intensity-scale. Large-scale influence on the duration exponent might lead to opposing trends for both ends of the duration range (Fig. \ref{fig:idv}c). Dependence of the intensity offset $\tau$ will mostly effect the long-duration regime (not shown).

\begin{figure}[h]%
\centering
\includegraphics[width=1\textwidth]{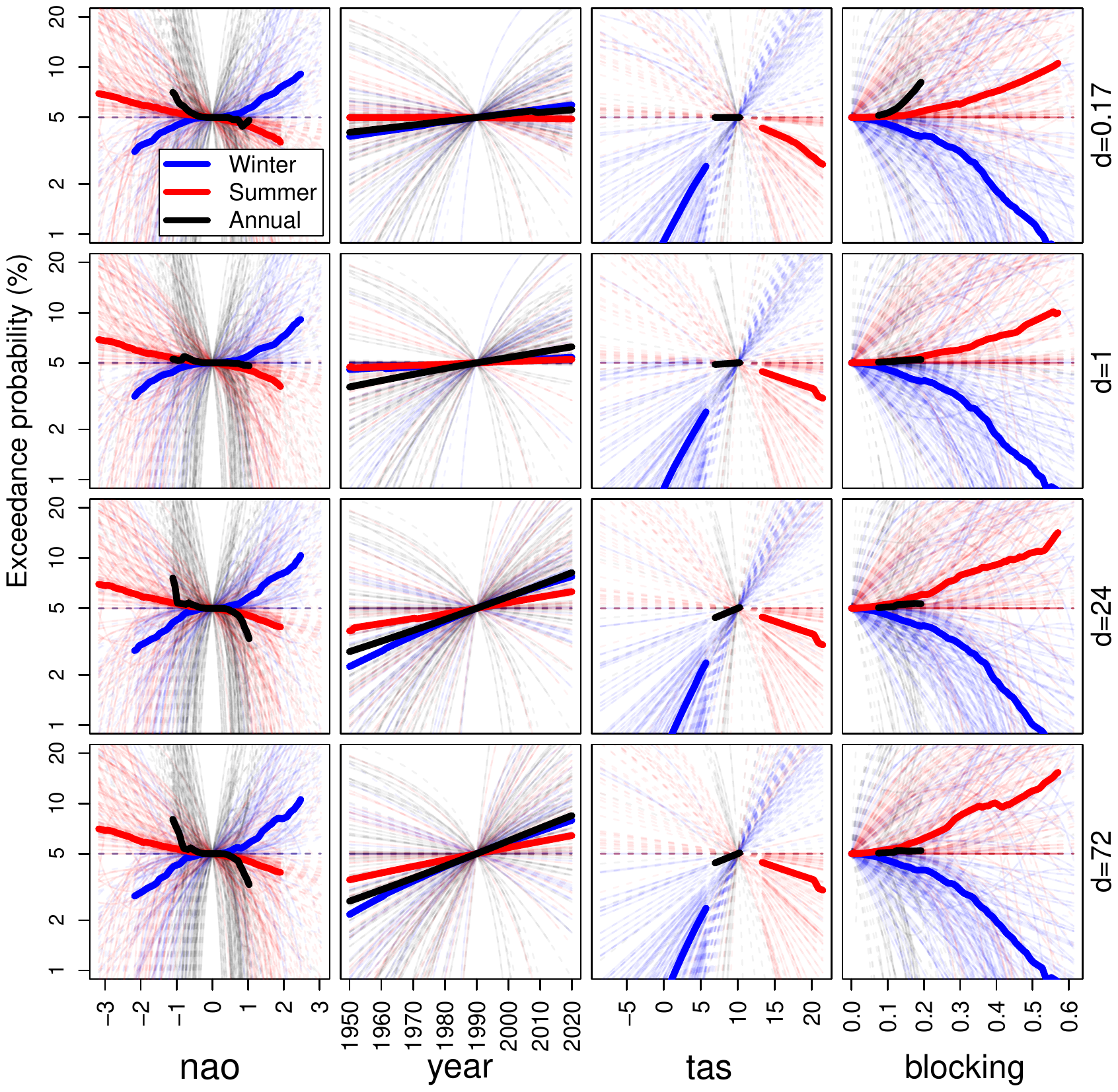}
\caption{Change of exceedance probability for an extreme event. An artificial extreme event was created as reference event. Its probability is $p_e=0.05$ in a situation which is defined as large-scale reference (see text). All curves meet at this value. Varying one large-scale variable (column) while fixing the other values allows to analyse the new exceedance probability $p_e$ of the reference event in the new large-scale situation. This figure shows how the probability of the reference event changes for different seasons (color) and durations (rows). Thin lines in the background represent individual stations/models while thick lines represent the median probability over all stations. Interpolations (extrapolations) are indicated as solid (dotted) thin lines for each station/model.}\label{fig:vf_all}
\end{figure}

However, the visualization of the dependence of extreme precipitation on large-scale variables is better done with a different plot design (Fig. \ref{fig:vf_all}). For the purpose of informing about large scale influence, the plot should show how extreme precipitation changes with respect to the large-scale variable on the x-axis and for many stations at once to improve robustness. In Fig. \ref{fig:vf_all}, a reference event was defined which has an annual exceedance probability $p_e$ (average return period) of 5\% (20 years) and the large scale reference parameter values for NAO $N$, year $y$, temperature $T$ (in $^{\circ}C$) and blocking $b$ are $N=0, y=1990, T=10, b=0$. Note, that all curves meet at these variable values. For each model (thin lines), one parameter (column) has been varied while all other parameters and the exceedance probability $p_e$ stay fixed to the large-scale reference values. Extrapolations outside of the data range of this parameter at this station are indicated as dotted lines. The thick lines show the median over all summer (red), winter (blue) and annual (black) models. In most cases, there is no difference between the durations (rows in Fig. \ref{fig:vf_all}, 1 min to 3 days, $d$ in hours). Only two clear duration-sensitive effects have been found: (1) The steepness of change with year increases for larger durations and (2) the annual models show a positive correlation between blocking index and probability of an extreme event only for minutely durations. The following results are similar for all durations. There is a positive correlation between NAO and probability of an extreme event in winter and a negative correlation in summer. The trend over time (year) is almost always positive, but smaller for short durations and smaller or zero in summer. Rising temperature has a positive effect on the probability of extreme rainfall in winter and a negative effect in summer. Blocking situations support extreme rainfall in summer and counteract extremes in winter.

\section{Summary and Discussion} \label{sec:discussion}
 
The aim of this study was to find meaningful large-scale variables that have an influence on extreme precipitation. Therefore, a parametrical duration-dependent GEV model includes large-scale parameters and non-stationarity. A stepwise BIC model selection was conducted and the results were verified with a cross-validated QSI. This way, IDF-curves can be produced, depending on large-scale variables. Furthermore, the influence of large-scale effects on extreme precipitation can be investigated. It was found that time has a positive correlation with probability of extremes for durations longer than 1 hour while the correlations of probability with nao, tas or blocking depend on the season. Especially the blocking situation, the NAO index and the temperature are covariates that can change the probability of an extreme event by a factor of 2 or more. 

Some of the results that are summarized in Fig. \ref{fig:vf_all} are unexpected.
The negative correlation between temperature and the probability of extremes (Fig. \ref{fig:vf_all}, third column) in summer is counter-intuitive. Therefore, we built a simple linear model ($RR_{\texttt{max}} \sim$ tas) for different durations and seasons (coefficients shown in Appendix Fig. \ref{fig:tas_trend}) and got the following results: In winter, the correlation is almost always positive. In annual models, the correlation shows no clear sign or is equally often positive and negative. in summer models, the correlation is negative for long durations (3 days) and slightly positive for short durations (1 minute), meaning that the intensity of extreme long-lasting rain events in summer decreases in case of warmer temperatures and the intensity of extreme convective events in summer increases slightly. To a certain extent, this is in accordance with the IPCC AR6 where the long-duration intensity shows a negative trend over time \citep{ipccAtlas_a, ipccAtlas_b}. Our complex large-scale d-GEV model (Fig. \ref{fig:vf_all}) follows the simple linear model (Fig. \ref{fig:tas_trend}) in most aspects. In both models, the correlation is almost always positive in winter and annual models. However, in summer only for short durations, the complex model differs and has a negative correlation with temperature. Short-duration results are supported by shorter time series of data than long-duration results, which could explain higher uncertainty and sensitivity to choice of model.

QSI values are in some cases not better for the complex large-scale model than for the model without large-scale information (blue regions in Fig \ref{fig:verification}), meaning that estimated quantiles are not necessarily modeled better than in the simpler model. However, the complex model is able to describe the influence of large-scale variables on extreme precipitation and provides new information and therefore has an advantage over the simple model, despite no better score in some cases. Furthermore, the fact that large-scale variables increased the BIC during the model selection process shows that the model profits from the use of these variables, since the BIC includes a penalty-term, which prevents overfitting. Still, the heterogeneous character of the out-of-sample-performance from the cross-validated QSI verification (Fig. \ref{fig:verification}) is noteworthy.

Large scale influence only marginally depends on the duration (Fig \ref{fig:vf_all}). But, using durations not only provides information, but also improves efficiency of data usage \citep{ulrich2020}. Therefore, it makes sense to use a duration-dependent model, however the duration-dependence does not play a large role in large-scale dependence.
  
When comparing our results with \cite{casanueva2014}, we find that both studies find the same opposite association with NAO in winter and summer over Germany. \citet[][Fig. 12]{lenggenhager2019} found an increase of precipitation with blocking defined over a European sector (0$^{\circ}$-30$^{\circ}$W) in summer. In winter, the chance of precipitation is decreasing. Both these findings are in accordance with our results. 

The aim of this study was to investigate the dependence of precipitation extremes on large-scale variables. There was no particular focus on the physical dynamics, leading to precipitation extremes. That is why the independent variables (NAO, temperature, blocking) were used in a large scale setting on purpose with no finer than monthly resolution. In future studies, we plan to investigate different scales like daily temperature or blocking and include seasonality in the model.

\section*{Declarations}
\subsection*{Acknowlegements}
We would like to express our gratitude to Thomas Junghänel for providing high-resolution data with long very time ranges for three stations (Köln-Bonn, Kall-Sistig, Nürburg-Barweiler). Furtermore, we say thank you to the DWD and to Marc Scheibel from the Wupperverband for maintaining and providing station-based data.

\subsection*{Funding}
This study is part of the \textit{ClimXtreme} project (Grant number 01LP1902H) and is sponsored by the Federal Ministry of Education and Research.

This work used resources of Deutsches Klimarechenzentrum (DKRZ) granted by its Scientific Steering Committee (WLA) under project IDs bb1152 and bm1159.

\subsection*{Data availability}
The annual maxima of rainfall and meta information of the measurement stations are available online \citep{data2022}.

\subsection*{Code availability} 
The IDF model can be used with our R-package, available via CRAN \citep{rIDF}.

\subsection*{Authors' contributions}
Conceptualization: Felix S. Fauer, Henning W. Rust; Methodology: Felix S. Fauer; Formal analysis and investigation: Felix S. Fauer; Writing - original draft preparation: Felix S. Fauer; Writing - review and editing: Felix S. Fauer, Henning W. Rust; Funding acquisition: Henning W. Rust; Supervision: Henning W. Rust. All authors read and approved the final manuscript.

\section*{Appendix}

\subsection{Model selection results for selected stations}\label{cha:apx_ms}

Table \ref{tab1:parameters} shows the d-GEV parameters for stations where at least 30 years of sub-hourly data are available. Stationary parameters are combined in the vector $\phi=\{\tilde{\mu},\sigma_0,\xi,\theta,\eta,\eta_2,\tau\}$ for summer and annual models or $\phi_w=\{\tilde{\mu},\sigma_0,\xi,\theta,\eta,\eta_2\}$ for winter, respectively. The other parameters show their functional dependency on large-scale variables in brackets, e.g., the shape parameter $\xi$ depending on time $t$ with a polynomial of third order notated as $\xi(t^3)$.

\begin{table}[h]
\begin{center}
\begin{minipage}{260pt}
\caption{d-GEV model parameters for chosen stations. The variable $\phi_w$ contains all parameters without large-scale dependence for winter and $\phi$ for summer and annual models. The large-scale variables are nao $N$, year $y$, tas $T$, and blocking $b$ with their corresponding polynomials.}\label{tab1:parameters}%
\begin{tabular}{@{}lll@{}}
\toprule
Station & Season & Dependencies\\
\midrule
 
Buchenhofen   &   DJF   &   $ \phi_w; \tilde{\mu}(t^2); \tilde{\mu}(b^1); \theta(y^3); \eta_2(N^1) $ \\   
          &   JJA   &   $ \phi; \tilde{\mu}(N^1); \sigma_0(N^4); \theta(y^1); \eta_2(b^2); \tau(t^1) $ \\   
          &   13   &   $ \phi; \xi(N^3); \theta(y^4); \eta_2(N^4) $ \\ \hline  
Leverkusen   &   DJF   &   $ \phi_w; \sigma_0(y^1); \sigma_0(t^2) $ \\   
          &   JJA   &   $ \phi; \sigma_0(y^4); \xi(N^2); \theta(y^1); \eta(y^4); \eta(b^1) $ \\   
          &   13   &   $ \phi; \sigma_0(y^4); \sigma_0(b^4); \theta(y^4); \eta(y^4) $ \\ \hline  
Neumühle   &   DJF   &   $ \phi_w; \sigma_0(t^3); \sigma_0(y^1) $ \\   
          &   JJA   &   $ \phi; \tilde{\mu}(N^1); \sigma_0(t^3); \tau(y^2) $ \\   
          &   13   &   $ \phi; \tilde{\mu}(t^4); \tilde{\mu}(b^4); \theta(b^4); \eta_2(b^1) $ \\ \hline  
Solingen-Hohenscheid   &   DJF   &   $ \phi_w; \tilde{\mu}(y^2); \sigma_0(t^3); \xi(y^3); \eta(N^4) $ \\   
          &   JJA   &   $ \phi; \tilde{\mu}(N^1); \sigma_0(y^4); \theta(y^2); \tau(t^1) $ \\   
          &   13   &   $ \phi; \theta(y^1) $ \\ \hline  
Seehausen                                  &   DJF   &   $ \phi_w; \tilde{\mu}(t^1) $ \\   
          &   JJA   &   $ \phi; \tilde{\mu}(b^2); \sigma_0(N^1); \theta(y^1) $ \\   
          &   13   &   $ \phi; \tilde{\mu}(y^1); \xi(y^1); \tau(N^1) $ \\ \hline  
Stötten   &   JJA   &   $ \phi; \tilde{\mu}(N^1); \sigma_0(N^2); \xi(y^4); \theta(N^1); \tau(y^3) $ \\   
          &   13   &   $ \phi; \tilde{\mu}(b^2); \tilde{\mu}(y^2); \sigma_0(b^1); \theta(t^1); \eta_2(b^1); \eta_2(t^1) $ \\ \hline  
Stuttgart-Echterdingen                     &   DJF   &   $ \phi_w; \sigma_0(t^4); \sigma_0(b^1); \xi(t^4) $ \\   
          &   JJA   &   $ \phi; \tilde{\mu}(N^1); \sigma_0(b^1); \theta(t^2); \eta(y^3) $ \\   
          &   13   &   $ \phi; \sigma_0(b^1); \theta(y^4); \tau(y^1) $ \\ 

\end{tabular}
\end{minipage}
\end{center}
\end{table}

\subsection*{Analysis of coefficients of a simple regression model}\label{secA2}

\begin{figure}[h]%
\centering
\includegraphics[width=10.2cm]{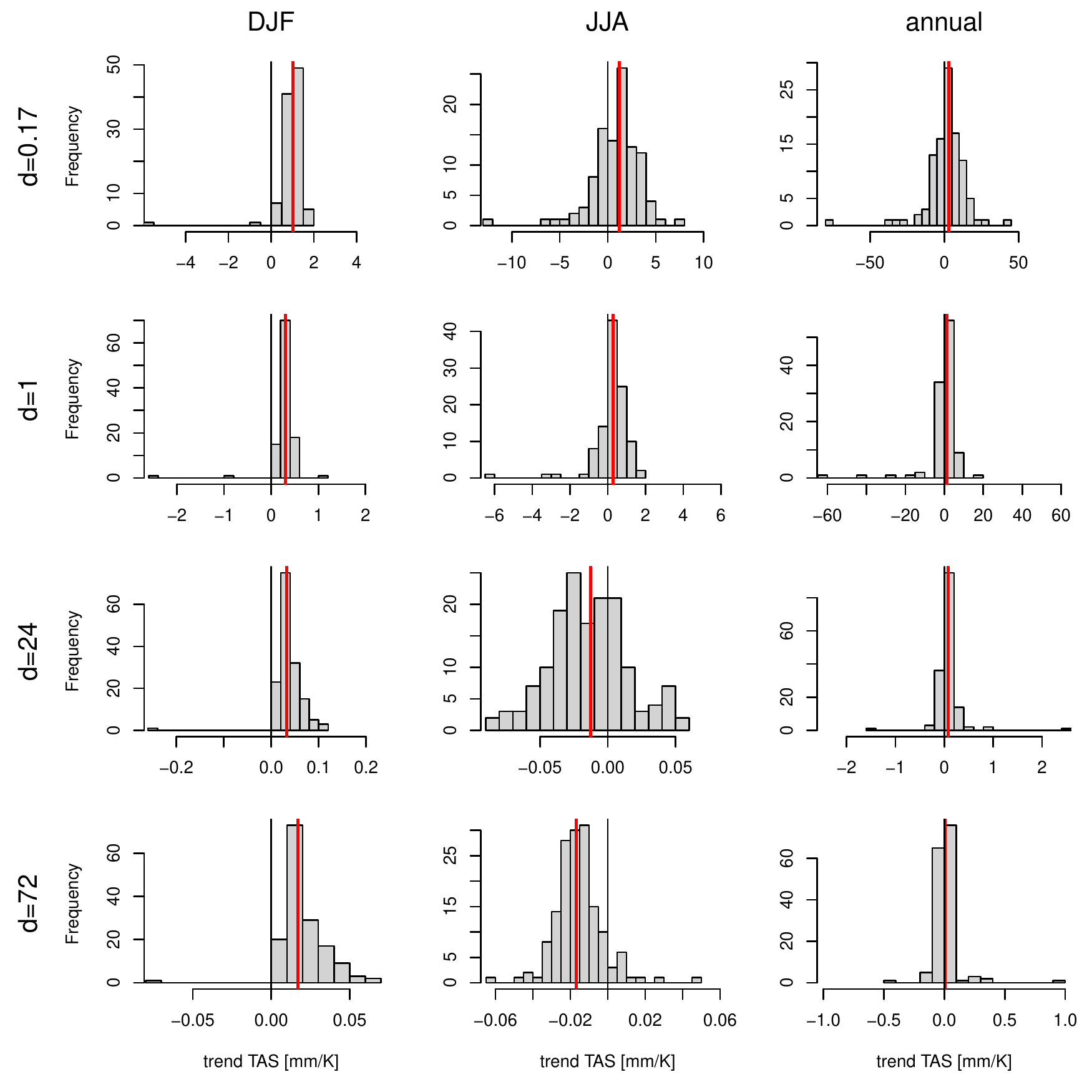}
\caption{Trend coefficients of temperature in linear model $RR_{\texttt{max}} \sim$temperature. The histogram shows the empirical distribution of coefficients over all stations. The vertical red line indicates the median.}\label{fig:tas_trend}
\end{figure}

To investigate the counter-intuitive result of negative correlation between summer temperature (TAS) and extreme precipitation, a simple model ($RR_{\texttt{max}} \sim$ tas) has been created. The coefficients (slope) of the parameter TAS are plotted in histograms in Fig. \ref{fig:tas_trend}. In winter (left column), most of the coefficients are positive as well as the median (red line). Coefficients for annual models (right column) are neither clearly positive or negative but centered around a slope of zero. These findings are in alignment with results from the complex non-stationary model (Fig. \ref{fig:vf_all}). Coefficients in summer (middle row) are rather positive for short and negative for long durations. Only for long durations (negative influence of TAS), the trend of the non-stationary model is supported. For short durations, both results seem contradictory, but effects of other large-scale variables in the complex model might explain the different correlation between TAS and $RR_{\texttt{max}}$ in summer for short durations.

\bibliographystyle{apalike}  

\bibliography{literature}

\end{document}